\documentclass{desyproc}

\begin{document}
\title{Search for the Dark Photon with the PADME Experiment at LNF}

\author{{\slshape Viviana Scherini$^1$ on behalf of the PADME Collaboration\footnote{
The PADME Collaboration is: P. Creti,  G. Chiodini, F. Oliva, V. Scherini (INFN Lecce);
A.P. Caricato, M. Martino, G. Maruccio, A. Monteduro, S.Spagnolo (INFN Lecce e Dip. di Matematica e Fisica, Universit\`a del Salento); P. Albicocco, R. Bedogni, B. Buonomo, F. Bossi, R. De Sangro,  G. Finocchiaro, L.G. Foggetta, A. Ghigo, P. Gianotti, M. Palutan, G. Piperno, I. Sarra, B. Sciascia, T. Spadaro, E. Spiriti (INFN Laboratori Nazionali di Frascati); C. Taruggi (INFN Laboratori Nazionali di Frascati and Universit\`a degli Studi di Roma "Tor Vergata"); L. Tsankov, (University of Sofia "St. Kl. Ohridski”); G. Georgiev, V. Kozhuharov (University of Sofia ``St. Kl. Ohridski'' and INFN Laboratori Nazionali di Frascati); F. Ameli, F. Ferrarotto, E. Leonardi, P. Valente (INFN Roma1); S. Fiore (INFN Roma1 e ENEA)
G.C.Organtini, M.Raggi (INFN Roma1 e Dip. di Fisica, ``Sapienza'' Universit\`a di Roma)}}\\[1ex]
$^1 $Istituto Nazionale di Fisica Nucleare, Sezione di Lecce, via per Arnesano I-73100 Lecce, 
Italy}

\contribID{familyname\_firstname}

\confID{16884}  
\desyproc{DESY-PROC-2017-XX}
\acronym{Patras 2017} 

\maketitle

\begin{abstract}

Massive photon-like particles are predicted in many extensions of the Standard Model 
with a hidden sector where dark matter is secluded. They are vector bosons mediating 
the interaction between dark matter particles and can be produced in scattering of
ordinary particles through a faint mixing to the photon. Most of the present experimental 
constraints on this ``dark photon'' (A') rely on the hypothesis of dominant decays to 
lepton pairs. The PADME experiment will search for the e$^+$e$^- \rightarrow \gamma $A' 
process in a positron-on-target experiment, assuming a decay of the A' into invisible 
particles of the hidden sector. The positron beam of the DA$\Phi$NE Beam-Test Facility, 
at Laboratori Nazionali di Frascati of INFN, will be used. A fine-grained, high-resolution 
calorimeter will measure the momentum of the photon in events with no other activity in the 
detector, thus allowing to measure the A' mass as the missing mass in the final state.
In about one year of data taking, a sensitivity on the interaction strength 
($\varepsilon^2$ parameter) down to $10^{-6}$ is achievable in the mass region M$_{\text{A'}}<23.7$~MeV.
The experiment is currently under construction and it is planned to take data in 2018. 
The status of PADME and its physics potential will be reviewed.

\end{abstract}

\section{Introduction}

The PADME experiment~\cite{PADMEex}, hosted in the DA$\Phi$NE~\cite{DAPHNE} Beam-Test 
Facility (BTF)~\cite{BTF} at Laboratori Nazionali di Frascati (LNF) of INFN, is designed 
to search for the dark photon by using an intense positron beam hitting a light target. 
The A' can be observed by searching for an anomalous peak in the spectrum of the missing mass 
measured in events with a single photon in the final state. The measurement requires the 
precise determination of the 4-momentum of the recoil photon, performed by an homogeneous 
electromagnetic calorimeter.

The collaboration aims to complete the design and construction of the experiment by the end 
of 2017 and to collect $\sim 10^{13}$ positrons on target by the end of 2018.

A comprehensive review of the dark sector, along with a summary of the ongoing 
experimental programs addressing the theoretical hypotheses motivating the search for the dark 
photon, can be found in Ref.~\cite{DARKS}.

\section{The PADME Detector}

\begin{figure}[ht!]
\centerline{\includegraphics[width=0.80\textwidth]{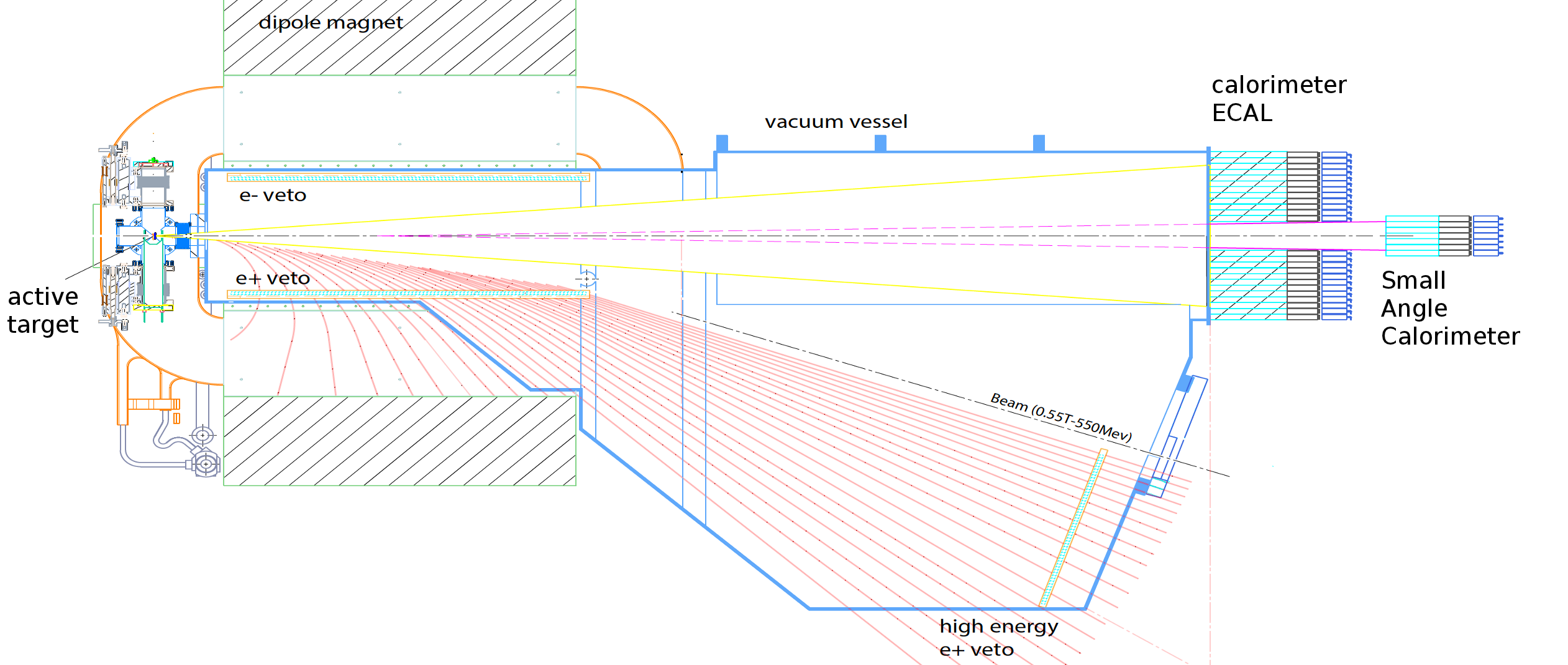}}
\caption{Layout of the PADME detector: the active target, the veto detectors 
inside the dipole magnet, the HEPVeto near to the non-interacting beam exit, the ECAL and the 
SAC.}\label{Fig:padmeEX}
\end{figure}

In its baseline configuration, PADME will use the positron beam of 550 MeV energy delivered 
by the LINAC of the DA$\Phi$NE. The time structure of the pulsed beam is a sequence of bunches 
with a controllable positron population (from single particle to about $10^4$) of constant 
intensity in a time span of 40 ns and a maximum repetition rate of 50 Hz. 
The experimental apparatus, whose layout is shown in Fig.~\ref{Fig:padmeEX}, consists of: 

\begin{itemize} 
\item{a $2 \times 2$ cm$^2$ polycrystalline CVD diamond active target~\cite{target,target17}, 
  aimed at measuring the beam intensity and position (with a precision of a few mm) by means of 
  perpendicular conductive strips. The low Z helps to reduce the occurrence of Bremsstrahlung 
  processes. The small thickness (100 $\mu$m) reduces the probability of e$^+$ multiple 
  interactions.}

\item{a dipole magnet, located 20 cm down-stream of the target, designed to deflect non-interacting 
  beam particles out of the detector and to direct the positrons that lost part of their energy 
  towards veto detectors. The field is 0.5 Tesla over a gap of 23 cm for 
  1 m length.}

\item{a system of  $1 \times 1 \times 18 $ cm$^3$ bars of plastic scintillators used as veto 
  detector for charged particles divided in 3 parts: two arrays of scintillator fingers on the 
  left and right internal walls of the dipole act as positron and electron veto, and another 
  one near the beam dump detects high energy positrons having lost a small part of their energy 
  (mainly through Bremsstrahlung processes).}

\item{a highly segmented electromagnetic calorimeter of cylindrical shape (ECAL)~\cite{RaggiECAL} 
  with axis on the beam line located at about 3 m from the target. Its final design implements 
  an active volume of 616 $2 \times 2 \times 23$ cm$^3$ BGO crystals covering the angular region 
  $20\div83$ mrad. The expected energy resolution is $\sim \frac{(1\div2)\%}{\sqrt{E}}$ for 
  $< 1$~GeV electrons and photons.}

\item{a Small Angle fast Calorimeter (SAC) made of $49 2 \times 2 \times 20 $cm$^3$ 
  lead glass (SF57) bars with angular coverage $0\div20$ mrad, placed behind the central 
  hole of the main calorimeter, thus instrumenting the region of maximum flux of Bremsstrahlung 
  photons produced in the target, mainly aimed at suppressing background from $3\,\gamma$ events.}
\end{itemize}

The target and all veto detectors will be hosted in a vacuum chamber to minimize the interactions 
of the beam with the atmosphere.

\section{Signal and Background}

The detector will identify events with a single photon generated in the e$^+$e$^-$ annihilation 
taking place in the interaction of the positron beam with the target. The dominant Standard Model 
processes expected to occur are Bremsstrahlung and e$^+$ e$^-$ $\rightarrow \gamma\gamma(\gamma)$. 
The probability that their kinematics will mimic a dark photon production event in the PADME 
detector can be reduced through an optimization of the ECAL geometry and granularity and of 
the veto system. The thin target and the adjustable beam intensity play a crucial role in 
reducing events pile-up.

A signal event is satisfying the following requirements: 
one cluster in the ECAL fiducial volume (with energy in a range optimized depending on 
M$_{\text{A'}}$), no hits in the vetoes, and no photons with energy larger than 50 MeV in the SAC.

\begin{figure}[hbt]
\centerline{\includegraphics[width=0.60\textwidth]{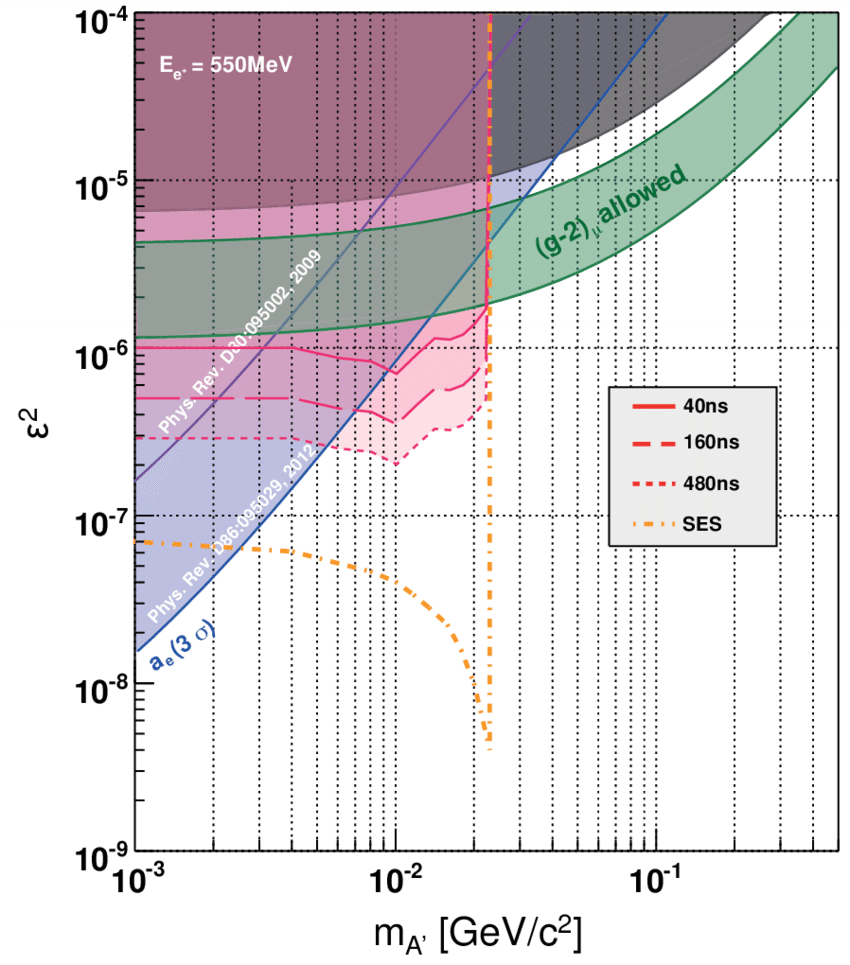}}
\caption{PADME estimated sensitivity for A' decaying into invisible particles for different values 
of the bunch length. The SES curve refers to single event sensitivity (no background).}\label{Fig:padmeINV}
\end{figure}

The sensitivity estimation is based on GEANT4 simulations extrapolated to $10^{13}$ e$^+$ 
positrons on target. This number of particles can be obtained by running PADME for 2 years at 
50\% efficiency with 5000 e$^+$ per 40 ns bunch at a repetition rate of 49 Hz. 
The obtained result for A' decaying into invisible particles is shown in Fig.~\ref{Fig:padmeINV}. 
Smaller values of the coupling constant $\epsilon$ can be explored by increasing the bunch length. 
The favored (g-2)$_\mu$ region can be explored in a model independent way (the only hypothesis on 
the A' is the coupling to leptons) up to masses of 23.7 MeV~\cite{RaggiECAL}. 
 
\section{Status and Perspectives}

The PADME experiment is pioneering for the first time the missing mass technique to constrain 
directly the A' invisible decay in the parameter region preferred by the (g-2)$_\mu$ and also 
to investigate other phenomena, not mentioned here, such as Axion Like Particles (ALPs)~\cite{alps},
Dark Higgs~\cite{dh} and the fifth force~\cite{5th}.
Early Monte Carlo studies applied to dark photon invisible decay modes demonstrated that PADME can 
reach a sensitivity down to the level of $\epsilon^2 \sim 1\cdot10^{-6}$ in the mass range 
M$_{\text{A'}}<23.7$~MeV.

The PADME experiment is expected to run in early 2018 and the preparation of the several detector 
components proceeds according to the schedule. In particular, the dipole magnet is ready and 
only the mechanical support for final integration must be prepared. In addition, the prototypes 
of calorimeters and charged veto detector systems have been finalized and tested. All the crystals 
and scintillator bars are ready for final assembly. A prototype diamond detector has been successfully 
tested and the final active diamond target is under construction. The readout and digitizing 
system ($1 \div 5$ Gs/s and 12bit ADC for about 1000 channels) is also available.

An upgraded pulsing system has been recently commissioned, allowing to deliver beam pulses of 
increasing length, up to $5\,\mu$s, so that, after having optimized the RF power and phases 
and the magnetic focusing in the LINAC, electron or positron beams could get accelerated 
close to the maximum energy, with a width of several hundreds of ns. An upgrade of the DA$\Phi$NE 
LINAC energy up to 1 GeV has been also proposed~\cite{BTF}. This will extend the parameter space 
to lower values of $\epsilon$ and increase the A' mass range to M$_{\text{A'}}<32$~MeV.


\begin{footnotesize}

\end{footnotesize}


\end{document}